\documentclass[english]{article}
\usepackage{fancyhdr}
\usepackage{color}
\usepackage{textcomp}
\usepackage{amsmath}
\usepackage{graphicx}

\makeatletter

\providecommand{\tabularnewline}{\\}

\newcommand{\lyxaddress}[1]{
	\par {\raggedright #1
	\vspace{1.4em}
	\noindent\par}
}

\makeatletter
\let\ps@plain\ps@fancy   
\makeatother

\makeatother

\usepackage{babel}
\begin{document}
\title{A/B Testing Measurement Framework for Recommendation Models Based
on Expected Revenue}
\author{Meisam Hejazi nia, Majid Hosseini, Bryant Sih }
\maketitle

\lyxaddress{\begin{center}
\textcolor{black}{Sparx -A Staples Digital Solution Innovation Lab,
West 5th Avenue, 3rd Floor, San Mateo, California 94402}\textcolor{red}{{} }
\par\end{center}}

\section{Abstract}

We provide a method to determine whether a new recommendation system
improves the revenue per visit ($RPV$) compared to the status quo.
We achieve our goal by splitting $RPV$ into conversion rate and average
order value ($AOV$). We use the two-part test suggested by Lachenbruch
to determine if the data generating process in the new system is different.
In cases that this test does not give us a definitive answer about
the change in $RPV$, we proposed two alternative tests to determine
if $RPV$ has changed. Both of these tests rely on the assumption
that non-zero purchase values follow a log-normal distribution. We
empirically validated this assumption using data collected at different
points in time from Staples.com. On average, our method needs a smaller
sample size than other methods. Furthermore, it does not require any
subjective outlier removal. Finally, it characterises the uncertainty
around $RPV$ by providing a confidence interval. 

\section{Introduction}

Measurement is a never-ending journey. Every measurement method has
its own drawbacks, so there is no one size fits all approach. Because
for retail recommendations the vast majority of outcomes (usually
> 95\%) are ``no-purchase'' events, the conversion rate is small
(large clumping at no-purchases). Furthermore, there are always customers
with large purchases, and thus, the distribution of non-zero order
values is not normal. Using Normal theory tests in this case causes
bias in estimates (\cite{lachenbruch1976analysis}). In addition,
outlier removal is a difficult task and in many cases subjective removal
might bias estimates. For example, when data is skewed and non-normal
(like the distribution of prices) studies show that sample size needs
to increase (\cite{muthen2002use}). Or in the case of non-normality,
bootstrap estimates might be inconsistent (\cite{andrews2000inconsistency,sen2010inconsistency}).
Finally, estimates that are efficient for clean data from simple distributions
(e.g. normal distribution which is symmetric and unimodal with thin
tail) may not be robust to contamination by outliers, and may be inefficient
for more complicated distributions, raising the need for very large
sample sizes for valid inference (\cite{Wikipedia2010}). 

One performance metric of a recommendation model is Revenue Per Visit
($RPV$). The relationship between $RPV$, Average Order Value ($AOV$),
and conversion rate r is as follows:
\begin{center}
\begin{equation}
RPV=r.AOV.
\end{equation}
\par\end{center}

To determine whether a new recommendation model has changed either
of $r$ or $AOV$, we can use a powerful test proposed by \cite{lachenbruch1976analysis}.
This test combines statistics for detecting differences in conversion
rate and $AOV$ to obtain a Chi-square valued measure that determines
if there is a statistically significant difference between the two
models. This test is powerful because it requires a small sample size.
We will describe Lachenbruch's test in the next section. In a general
sense, comparisons can be defined in 3 exclusive states: significantly
positive (Pos. Sig.), non-significant (Non-Sig.), or significantly
negative (Neg. Sig.). Since the objective of a recommendation system
is to increase $RPV$, we can visualize the outcomes of a significance
test of $r$ and $AOV$ on $RPV$ based on Table 1, where + is a guaranteed
significantly positive difference in $RPV$, \textminus{} is a decrease,
0 is no difference, and X is indeterminate. For most of the cases,
Lachenbruch's test tells us with a relatively small sample size whether
the new recommendation system is more effective. For the two indeterminate
cases we develop two parametric tests of significance. 

\begin{table}

\caption{Expected Revenue Significance Test}

\begin{centering}
\begin{tabular}{|c|c|c|c|}
\hline 
 & $AOV$ Pos. Sig. & $AOV$ Non-Sig.  & $AOV$ Neg. Sig.\tabularnewline
\hline 
\hline 
$r$ Pos. Sig. & + & + & X\tabularnewline
\hline 
$r$ Non-Sig. & + & 0 & -\tabularnewline
\hline 
$r$ Neg. Sig. & X & - & -\tabularnewline
\hline 
\end{tabular}
\par\end{centering}
\end{table}

\section{Tests of Significance }

In this section, we will describe in detail the process to determine
whether $RPV$ of the new recommendation is different from the old
recommendation system in a statistically significant sense. To summarize
this process, its stages are as follows: 

1. Run a pilot study and determine the sample size for Lachenbruch's
Test. 

2. Run Lachenbruch's Test and if it results in one of the determinate
cases, then stop the test with a definite conclusion (See Table 1). 

3. If Lachenbruch's Test is indeterminate, run a new pilot study to
determine the sample size for either of our newly devised parametric
tests (derived in the next section). 

4. Run the new parametric test, and conclude if there is a statistically
significant difference for $RPV$ between the two models.

\section{Lachenbruch\textquoteright s Two-Part Test }

Lachenbruch's Test applies to data with clumping at zero. Clumping
at zero for recommendation systems can be interpreted as no purchase.
This test compares the distribution of $RPV$ across control and treatment
groups. In other words, the null hypothesis in this test asserts the
equality of these two distributions. To run this test, \cite{lachenbruch1976analysis}
uses two statistical measures: $z_{p}$ and $z_{U}$. The value $z_{p}$
denotes z-statistics for proportion of no purchases, and it has a
normal distribution in the limit, so its square has an asymptotic
Chi-square distribution with one degree of freedom. The value $z_{U}$
is a function of Mann-Whitney-Wilcox U statistic for order values
and has an asymptotic normal distribution, so its square has Chi-square
distribution with one degree of freedom. As a result, Lachenbruch's
statistic $L=z_{p}^{2}+z_{U}^{2}$ has an asymptotic Chi-square distribution
with two degrees of freedom. If $L$ is greater than 9.633 we can
reject the null hypothesis with 95\% confidence. The statistic $z_{p}$
is defined as follows: 

\begin{equation}
z_{p}=\frac{p_{1}-p_{2}}{\sqrt{p_{p}(1-p_{p})(\frac{1}{n_{1}}+\frac{1}{n_{2}})}}
\end{equation}
where $p_{1}$, $p_{2}$ denote the no-purchase probabilities for
the treatment and control groups, respectively. The value $p_{p}$
denotes the no purchase probability for the pooled group (i.e. $p_{p}=\frac{k_{1}+k_{2}}{n_{1}+n_{2}}).$
The values $k_{1}$, $k_{2}$ denote the observed number of no purchases
for the treatment and control groups, respectively. The values $n_{1}$,
$n_{2}$ denote the sample sizes for the treatment and control groups,
respectively.

The value $z_{U}$ is defined as follows:

\begin{equation}
z_{U}=\frac{U-\frac{n_{1}n_{2}}{2}}{\sqrt{n_{1}n_{2}(n_{1}+n_{2}+1)/12}}
\end{equation}
where $U$ is the Mann-Whitney-Wilcoxon statistic for order values.\footnote{Computing rank-sum $U$ involves counting the number of times the
first value wins over any observations in the other set (the other
value loses if the first is larger) with 0.5 for any ties. The sum
of wins and ties construct the $U$ measure. $U$ can be calculated
using most statistical packages or python (scipy.stats mannwhitneyu).}. 

If the Chi-square test suggests significant difference in the distributions
of $RPV$ between the treatment and control groups, \cite{lachenbruch1976analysis}
suggests we can compare the values of $z_{p}$ and $z_{U}$ against
the standard normal distribution (i.e. 1.96 and -1.96 as a threshold
for 95\% confidence) to identify the sources of the difference. Knowing
the sources of the difference between the two distributions, we can
use Table 1 in order to identify whether the expected revenue between
treatment and control groups are statistically different. In the cases
where Table 1 is indeterminate, we need to use another statistical
test. This test requires collecting a new data set (i.e. we can not
use the old data set for the new test). 

\begin{figure}
\caption{\label{fig:The-Two-Stage}The Two-Stage Model where the first stage
is Customer Decision to Purchase (conversion) and the second stage
is, if decision was to purchase, order value. Order value is often
modeled as a log-normal distribution \cite{Feldman2013}.}
\includegraphics[width=0.8\textwidth]{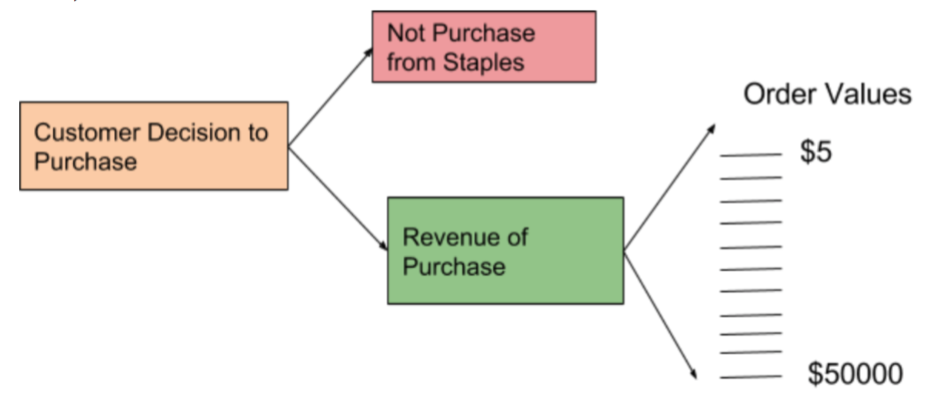}
\end{figure}

\section{Tests for Detecting Change In Expected Revenue }

In this section, we introduce two tests that assume a parametric form
for $RPV$ and attempt to establish if there is any change in $RPV$
between treatment and control groups. Both tests assume that the data
is generated in a two-stage process, which mixes non-purchases with
log-normally distributed order values (See Figure \ref{fig:The-Two-Stage}).
The first test uses the multivariate Delta method to estimate the
variance of the estimated expected revenue. The second test uses the
likelihood-ratio test to make inference. A log-normal distribution
is a good candidate to deal with fat tails and is widely used in science
and business (\cite{limpert2001log}). This distribution can arise
due to wealth and income distribution (\cite{gabaix2009power})
or the cognitive lock-in of customers to the Staples website (\cite{johnson2003cognitive}).
We checked Staples.com order values data for log-normal distribution
using $Q-Q$ plots, and found that it fits (See Appendix). 

\subsection*{Method 1: Expected Revenue Confidence Interval Using the Delta Method}

To compare expected revenues between treatment and control groups,
we will derive a confidence interval for this difference. This confidence
interval depends on our estimates for $r$ ($\hat{r}$) and $AOV$
($\hat{AOV}$), and our uncertainty around them. The estimates $\hat{r}$
and $\hat{AOV}$ are independent. Therefore, their covariance is zero.
As a result, we can use the multivariate Delta method\footnote{The Delta Method, Theorem 20 of http://www.stat.cmu.edu/\textasciitilde larry/=stat705/Lecture4.pdf,
accessed on Feb 21st, 2017.} to estimate the variance of $RPV$ from our point and variance estimates
of $\hat{r}$ and $\hat{AOV}$. Specifically: 

\begin{equation}
\sigma_{RPV}^{2}=\hat{AOV}^{2}\hat{\sigma}_{r}^{2}+\hat{r}^{2}\hat{\sigma}_{AOV}^{2}
\end{equation}
where all these values are pooled estimates (i.e. the estimates are
computed over a sample that consists of combined observations from
treatment and control groups). 

To estimate $AOV$ and our uncertainty around it from purchase values,
which we assumed to have a log-normal distribution, we transform the
purchase values by the log function. These transformed values have
a normal distribution, so we use its mean $\hat{\mu}$ as a measure
for its location, and its standard deviation $\hat{\sigma}_{AOV}$
as a measure for uncertainty around it. 

Since $\hat{r}$ is a binomial proportion, its estimated variance
is: 

\begin{equation}
\hat{\sigma}_{r}^{2}=\hat{r}(1-\hat{r})
\end{equation}
where $\hat{r}$ is the pooled proportions of non-purchases. 

As a result, we can write the estimated variance of $RPV$ as follows: 

\begin{equation}
\hat{\sigma}_{RPV}^{2}=e^{\hat{\mu}+\hat{\sigma}_{AOV}^{2}/2}\hat{r}(1-\hat{r})+\hat{r}^{2}(e^{\hat{\sigma}_{AOV}^{2}}-1)e^{2\hat{\mu}+\hat{\sigma}_{AOV}^{2}}
\end{equation}

To make inference, the final step is computing the convergence rate
for the asymptotic distribution, which is equal to the harmonic mean
of sample sizes: 

\begin{equation}
h(n_{1}+n_{2})=(\frac{1}{n_{1}}+\frac{1}{n_{2}})^{-1}
\end{equation}

As a result, the 95\% confidence interval for the expected revenue
$RPV$ can be computed as follows: 

\begin{equation}
(\hat{RPV}_{treatment}-\hat{RPV}_{control})\pm1.96\sqrt{\frac{\hat{\sigma}_{RPV}^{2}}{h(n_{1},n_{2})}}
\end{equation}

\subsection*{Method 2: Likelihood Ratio Test for Expected Revenue }

Likelihood ratio test allows comparison between treatment and control
groups parametrically. Based on Wilk\textquoteright s Theorem, the
log likelihood ratio has an asymptotic Chi-square distribution with
the degree-of-freedom (df) equal to the difference between degrees
of freedom of the null and alternative hypotheses. The likelihood
ratio test rejects the null hypothesis if the value of this statistic
is too small. The Neyman-Pearson lemma states that likelihood ratio
test is the most powerful test of significance. Many common test statistics
such as Z-test, F-test, Pearson\textquoteright s chi-square test,
and G-test can be phrased as log-likelihood ratio test or its approximation.
In order to define the likelihood ratio test for expected revenue,
we need to specify a likelihood function under the null and alternative
hypotheses. Then we form the likelihood ratio value which is defined
as follows:

\begin{equation}
LR=\frac{L(H_{0})}{L(H_{1})}
\end{equation}
where $L$ is the likelihood function. $LR$ is a number between zero
and one. The low values of the likelihood ratio mean that the observed
result was less likely to occur under the null hypothesis as compared
to the alternative. 

For our expected revenue test, null hypothesis likelihood function
is defined as follows: 

\begin{equation}
L(H_{0})=LN(OV_{c}|\mu_{c},\sigma_{c}^{2})Bin(n_{c},k_{c}|p_{c})LN(OV_{t}|\mu_{t},\sigma_{t}^{2})Bin(n_{t},k_{t}|p_{t})
\end{equation}
under
\[
p_{c}\mu_{c}=p_{t}\mu_{t},
\]
where $OV$ denotes the order value, $LN$ denotes log-normal distribution,
and $Bin$ denotes the binomial distribution. The null hypothesis
puts constraints over the parameter space of the likelihood function.
This constraint requires that the expected revenues under control
and treatment groups are equal. Similarly, but excluding the constraint,
we can define the likelihood function under the alternative hypothesis
as follows:

\begin{equation}
L(H_{1})=LN(OV_{c}|\mu_{c},\sigma_{c}^{2})Bin(n_{c},k_{c}|p_{c})LN(OV_{t}|\mu_{t},\sigma_{t}^{2})Bin(n_{t},k_{t}|p_{t})
\end{equation}

In order to run the likelihood ratio test, we need to first find the
maximum likelihood estimates of parameters (i.e. $\mu_{c},\sigma_{c}^{2},p_{c},\mu_{t},\sigma_{t}^{2},p_{t}$).
To find these parameters for the alternative hypothesis, we can use
the closed form solution we specified in the Delta method section
because we don\textquoteright t have any constraints. 

As maximizing the likelihood function is equivalent to maximizing
the log likelihood function, and derivatives of the log likelihood
functions are analytically more tractable than that of the likelihood
function, we use the log likelihood function rather than the likelihood
function. We define the Lagrangian as follows:

\begin{equation}
\begin{array}{c}
Log(H_{0})=\log LN(OV_{c}|\mu_{c},\sigma_{c}^{2})+\log Bin(n_{c},k_{c}|p_{c})+\log LN(OV_{t}|\mu_{t},\sigma_{t}^{2})\\
+\log Bin(n_{t},k_{t}|p_{t})-\lambda(p_{c}\mu_{c}-p_{t}\mu_{t})
\end{array}
\end{equation}

In order to optimize the Lagrangian, we compute its gradient with
respect to the parameter vector (i.e. $(\mu_{c},\sigma_{c}^{2},p_{c},\mu_{t},\sigma_{t}^{2},p_{t},\lambda)$)
and equate it to a vector of zeros. As a result, we will have the
following system of equations to solve for the parameter vector: 

\begin{equation}
\begin{cases}
\begin{array}{c}
k_{c}(1-p_{c})-p_{c}(n_{c}-k_{c})-\lambda\mu_{c}p_{c}(1-p_{c})=0\\
k_{t}(1-p_{t})-p_{c}(n_{t}-k_{t})-\lambda\mu_{t}p_{t}(1-p_{t})=0\\
\log AOV_{c}-\mu_{c}-\lambda\mu_{c}^{2}p_{c}=0\\
\log AOV_{t}-\mu_{t}-\lambda\mu_{t}^{2}p_{t}=0\\
\begin{array}{c}
(\log AOV_{c}-\mu_{c})^{2}=\sigma_{c}^{2}\\
(\log AOV_{t}-\mu_{t})^{2}=\sigma_{t}^{2}\\
\mu_{t}p_{t}-\mu_{c}p_{c}=0
\end{array}
\end{array}\end{cases}
\end{equation}

We did not find a closed form solution for this system of equations,
so we use a nonlinear optimization algorithm to numerically find these
parameters. Specifying search procedures that gives us this parameter
vector completes our specification of this test. 

\section{Sample Size Estimation}

In this section we determine the sample size required in order to
obtain the desired statistical power for each of the tests that we
discussed above. Note that although we carry two tests sequentially
(i.e. Lanchenbruch\textquoteright s Test followed by either likelihood
ratio or Delta method test), the overall sample size we will need
for these two tests is going to be smaller than average. This is because,
in many cases Lachenbruch\textquoteright s Two-part Test can establish
whether the distribution of active and control groups are the same
or not. 

The standard process in hypothesis testing consists of collecting
data, followed by the execution of the test. In the testing stage,
two possible errors can happen: rejecting the null hypothesis when
it is true (Type I error), and not rejecting the null hypothesis when
it is false (Type II error). In our case, null hypothesis states that
the treatment and control groups have the same distribution. As a
consequence, a Type I error will make us believe that we have an improvement
over our existing system, while in reality we do not. On the other
hand, a Type II error makes us believe that the new system is not
better than the old system, while in reality it is, so it costs us
the opportunity to replace the existing system with a better one. 

The probability of committing a Type I error is a function of the
shape of the null model distribution. The magnitude of a Type II error
is a function of the sample size and the model. Specifying the magnitude
of these two types of error allows us to find the sample size required
for the test. 

\subsection*{Sample size determination using Monte Carlo Method }

The Monte Carlo simulation method is applicable for all tests that
we discussed above. Steps of this process are as follows: 
\begin{enumerate}
\item Run a pilot study to estimate the magnitude of the effect and uncertainty
around it. 
\item Specify the desired statistical power level. 
\item Start with an initial guess about the required sample size. 
\item Repeat many times (e.g. 100 or 1000 times) the following: 
\begin{enumerate}
\item Sample with replacement from the data collected in the pilot study
as many times as the sample size specified in the previous step. 
\item Run the statistical test for each of the samples. 
\item Compute the proportion of times that this test identifies the effect
as significant. 
\item Consider this proportion as the power of the statistical test. 
\end{enumerate}
\item Adjust the sample size by comparing the power of the statistical test
with the desired power. If you need a higher power, increase the sample
size, and run step 2 again. 
\end{enumerate}

\section{Conclusion, Limitations, and Future Work}

In this paper, we addressed the problem of incremental revenue measurement
and valid statistical inference. Our approach has three distinct advantages.
First, it allows analysts to conclude whether a new system is effective
using a smaller sample size. Second, our approach does not require
subjective outlier removal. Third, our approach provides a method
to determine a confidence interval for $RPV$. 

Although our approach improves the current measurement methods, it
does not account for the dependence between multiple visits of a given
visitor. We address this issue in a forthcoming paper.

\section*{Acknowledgements}

We thank Karthik Kumar, Stephanie Whang, Ryan Applegate, and Nitin
Varma for their helpful discussions.

\bibliographystyle{acm}
\nocite{*}
\bibliography{ABTestingResultMetricsForRecommendationModelsBasedonOverallExpectedRevenue}

\section*{Appendix- Test of Log-Normality}

In order to empirically validate the log-normality of the distribution
of revenue per purchase, there are a variety of Normality tests including
Jarque-Bera, Kolmogorov-Smirnov, $Q-Q$ plot, and Shapiro-Wilk tests.
The latter test (i.e. Shapiro-Wilk test) might not be appropriate
for the era of big data\footnote{StackOverflow, Is normality test useless , http://stats.stackexchange.com/questions/2492/is-normality-testing-essentially-useless,
accessed on Feb 27, 2017.}. Other statistical tests such as Kolmogorov-Smirnov and Jarque-Bera
are also not appropriate because they raise a red flag with a small
deviation from normality, so unless the data has an exactly normal
distribution, they return a verdict that the data is not normal\footnote{Stackoverflow, Why would all the tests for normality reject the null
hypothesis, http://stats.stackexchange.com/questions/16611/why-would-all-the-tests-for-normality-reject-the-null-hypothesis,
accessed on Feb 27, 2017. }. As a result, we resort to the $Q-Q$ plot approach.

\textbf{$Q-Q$ }plot test: a quantile-quantile plot compares the cumulative
probability density function for two random variables (e.g. empirically-derived
distribution vs. known theoretical distribution). It plots the quantiles
of two distributions against each other, and in the case where the
points lie on the $y=x$ line, it suggests that the distribution of
empirical data and theoretical are the same. The advantage of a $Q-Q$
plot over other \textquotedblleft goodness of fit\textquotedblright{}
measures is that it is easily recognizable visually. To generate the
$Q-Q$ plot we first sort the empirical data into non-decreasing order
to yield an inverse probability density function. In addition, we
create a vector of indexes from 1 to the number of elements, and calling
each index $i$, we create a vector of $\frac{i-0.5}{n}$, where $n$
is the number of observations. We call this the vector of cumulative
probability. Next, we create a vector of $z$-scores based on the
cumulative probability vector (i.e. the corresponding normally distributed
random variable threshold, for which we will use to determine the
inverse of the normal distribution). Then we plot the sorted empirical
data versus $z$-scores. If there is no systematic deviation from
the equality line, then the distribution of empirical data is normal.
To test for log-normality, we need to first log transform and standardize
the empirical data (i.e. extract mean and divide by standard deviation)
and then generate the $Q-Q$ plot\footnote{Kratz, M., \& Resnick, S. I. (1996). The $Q-Q$-estimator and heavy
tails. Stochastic Models, 12(4), 699-724. }. 

\begin{figure}
\caption{A $Q-Q$ plot of log(Staples.com order prices) from the period of
Feb 17, 2017 to Feb 28, 2017 showing a log-normal distribution.}

\centering{}\includegraphics[width=0.8\textwidth]{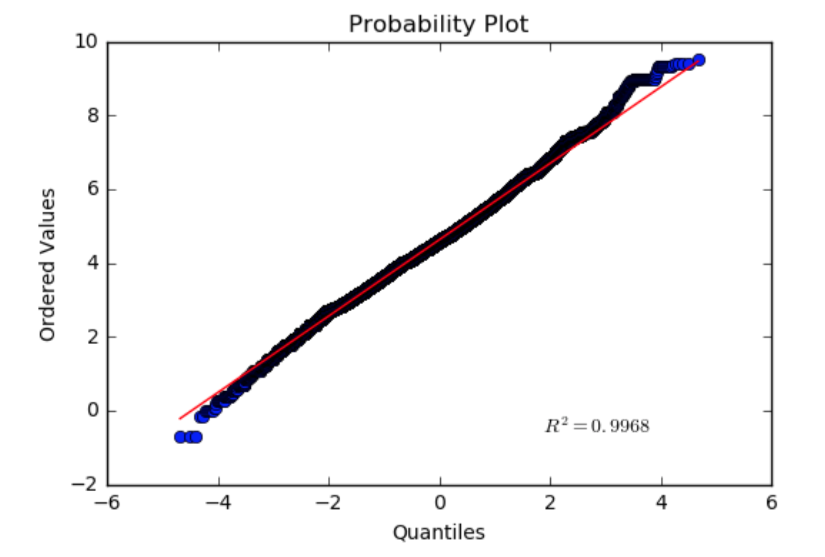}
\end{figure}

\end{document}